\documentclass[twocolumn,pra,showpacs,superscriptaddress,showpacs]{revtex4-1}
\usepackage{amssymb,amsmath}
\usepackage{graphicx}
\usepackage{bm}
\usepackage{dcolumn}
\usepackage{float}
\usepackage{url}
\usepackage{color}
\usepackage{subfigure}
\usepackage{txfontsb}
\usepackage[driverfallback=dvipdfm]{hyperref}
\hypersetup{pdfpagemode=FullScreen,colorlinks=true,breaklinks,urlcolor=blue,linkcolor=blue,citecolor=blue}

\usepackage{CJK}
\usepackage{slashed}
\usepackage{braket}

\begin{document}
\title{Stability of entanglement-spectrum crossing in quench dynamics of one dimensional gapped free-fermion systems}
\author{Shuangyuan Lu}
\affiliation{School of Physics, Peking University, Beijing 100871, China}
\affiliation{Institute for Advanced Study, Tsinghua University, Beijing, 100084, China}
\author{Jinlong Yu}
\email{jinlong.yu.physics@gmail.com}
\affiliation{Institute for Advanced Study, Tsinghua University, Beijing, 100084, China}
\affiliation{Center for Quantum Physics, University of Innsbruck, Innsbruck A-6020, Austria}
\affiliation{Institute for Quantum Optics and Quantum Information of the Austrian Academy of Sciences, A-6020 Innsbruck, Austria}
\date{\today}
\begin{abstract}
{In a recent work by Gong and Ueda (arXiv:1710.05289), the classification of (1+1)-dimensional quench dynamics for the ten Altland-Zirnbauer classes is achieved, and entanglement-spectrum crossings of the time-dependent states for the topological classes (AIII, DIII, CII, BDI, and D) are discovered as a consequence of the bulk-edge correspondence. We note that, their classification scheme relies on the limit that the energy spectrum of the post-quench Hamiltonian is flat, because any finite band dispersion leads to the break down of time-reversal and chiral symmetries for the parent Hamiltonian (which are used for the classification). We show that, because of the reduction of symmetry by finite energy dispersion, the gapless entanglement-spectrum crossing in the flat-band limit in classes AIII, DIII, and CII is unstable, and could be gapped without closing the bulk gap. The entanglement-spectrum crossing in classes BDI and D is still stable against energy dispersion. We show that the quench process for classes BDI and D can be understood as a $\mathbb{Z}_2$ fermion parity pump, and the entanglement-spectrum crossing for this case is protected by the  conservation of fermion parity.
}
\end{abstract}

\maketitle

%\tableofcontents

\section{Introduction}

The study of topological materials has been an active research field in the past a few decades due to the novel properties and potential applications in spintronics and quantum computation~\cite{Hasan2010Review,Qi2011Review}. The topological classification of the gapped free fermions (i.e., noninteracting topological insulators and mean-field topological superconductors) in all spatial dimensions for a total of ten Altland-Zirnbauer (AZ) classes~\cite{Altland-Zirnbauer1997} that based on the presence or absence of time-reversal, particle-hole, and chiral symmetries is also achieved~\cite{Schnyder-Ryu2008,Kitaev2009,Ryu2010}; for a review, see, e.g., Ref.~\cite{chiu2016classification}. According to the bulk-edge correspondence~\cite{Prodan2016Book}, the topological invariant defined for the bulk indicates the presence of gapless boundary states that are stable against symmetry-preserving perturbations as well as disorder~\cite{chiu2016classification}. As an alternative of setting a boundary in the real space and studying the gapless boundary states, stable gapless boundary modes of the entanglement spectrum (ES)~\cite{li2008entanglement} for a quantum state with an artificial cut in real space can also be used to characterize the topology of the system~\cite{fidkowski2010entanglement,Turner2010}. 

In addition to revealing the topology for equilibrium states, the study of topology for out-of-equilibrium systems has also attracted great interests, partially due to the rapid progress of several recent cold-atom experiments~\cite{Flaschner2018,Sengstock2017,USTC-PKU2018}. As first noted in Ref.~\cite{wang2017scheme}, for the quench process of a two-dimensional Chern insulator (e.g., the Haldane model~\cite{Haldane1988}, or the quantum anomalous Hall model~\cite{Qi2006}, as realized in cold-atom setups in Refs.~\cite{Jotzu2014} and \cite{USTC-SOC_2016}, respectively), if we prepare the initial state as a trivial state with zero Chern number~\cite{TKNN1982} and evolve the system with another Hamiltonian, it is possible to define a \emph{dynamical} Hopf invariant for the (2+1)-dimensional system albeit the Chern number of the system at any particular time is trivially zero~\cite{Rigol2015}. It is proved that~\cite{wang2017scheme}, for this quench process, the dynamical Hopf invariant equals the Chern number of the post-quench Hamiltonian, and thus the topological information of the Hamiltonian can be extracted from the dynamical evolution of a two-dimensional {\lq\lq}trivial state{\rq\rq}. The geometric essence of the Hopf invariant -- the linking number -- has been extracted experimentally for both the topologically trivial~\cite{Flaschner2018} and nontrivial~\cite{Sengstock2017,USTC-PKU2018} post-quench Hamiltonians. 

Parallel to the study of (2+1)-dimensional systems, the topology of (1+1)-dimensional quench dynamics is also studied recently~\cite{Yang-Chen2018,gong:2017,mcginley2018topology}. It is found that, a dynamical Chern number can be defined for certain (1+1)-dimensional quench processes~\cite{Yang-Chen2018,gong:2017}. Moreover, a complete classification of (1+1)-dimensional quench dynamics for the ten AZ classes is achieved for post-quench Hamiltonians with flat bands~\cite{gong:2017}, and stable ES crossing is observed~\cite{gong:2017} as a consequence of bulk-edge correspondence for the entanglement Hamiltonians~\cite{fidkowski2010entanglement}. The parent Hamiltonian that is used for the classification in Ref.~\cite{gong:2017} considers band-flattened post-quench Hamiltonian. The problem is that, when band dispersion of the post-quench Hamiltonian is taken into account, the parent Hamiltonian suffers from {\lq\lq}dynamically induced symmetry breaking{\rq\rq}~\cite{mcginley2018topology}, and the bulk topology that calls for symmetry protection becomes fragile against band dispersion. To be specific, time-reversal and chiral symmetries that relate time-dependent states/parent Hamiltonians between time $t$ and $-t$ are generally broken for dispersive post-quench Hamiltonians~\cite{mcginley2018topology}, and the number of topologically nontrivial classes is greatly suppressed. We show that, the remaining two topological classes are BDI and D that preserve the particle-hole symmetry. The stability of the ES crossing for these two classes can be understood as a parity pump~\cite{kitaev2001unpaired,Fu-Kane2009,teo2010topological}, and is protected by the conservation of fermion parity~\cite{teo2010topological}. For other classes (AIII, DIII, and CII) that are claimed to be topological in the flat-band classification scheme~\cite{gong:2017}, we show that the corresponding ES crossing is generally unstable against band dispersion of the post-quench Hamiltonian. 

This paper is organized as follows. In Sec. \ref{Sec:main_results}, we review the flat-band classification result for (1+1)-dimensional systems in Ref.~\cite{gong:2017}, and show the stability of the corresponding topological classes against band dispersion. In Sec. \ref{Sec:stable}, we analysis the stable classes D and BDI, and show that the corresponding stable single-particle ES crossing can be understood as a parity pump process. 
In Sec. \ref{Sec:unstable}, we show, through explicit calculation for the SSH-like model in class AIII, that the ES crossing in the flat-band case can be gapped by considering dispersive post-quench Hamiltonians. The unstable classes DIII and CII can also be understood through the results for class AIII. We conclude in Sec. \ref{Sec:conclusion}. 

\section{Main results}\label{Sec:main_results}

\begin{table*}[htbp]
 \caption{\label{Tab:classification}The classification of (1+1)-dimensional systems with flat-band Hamiltonians in the presence/absence of time reversal (T), particle-hole (C), and chiral (S) symmetries~\cite{gong:2017}, and its stability against band dispersion. As shown in \cite{gong:2017}, there are all together five topological classes realized by quench dynamics when treating flat-band Hamiltonians: AIII, BDI, D, DIII, and CII. We show that, when considering finite band dispersions, three of them (AIII, DIII, and CII) become topologically trivial. The topology of class D ($\mathbb{Z}_2$) remains intact, but the topology of class BDI is reduced from $\mathbb{Z}$ to $\mathbb{Z}_2$. The stability of the $(1+1)$-dimensional bulk topology is check by the stability of the entanglement-spectrum crossing according to the bulk-edge correspondence.}
 \begin{center}
 \begin{tabular}{c ccc cc}
  \hline\hline
  Altland-Zirnbauer class $\,\,\,$ & T & C & S & $\,\,\,$Flat-band classification for (1+1)D~\cite{gong:2017}$\,\,\,$ & Stable against band dispersion? \\
  \hline
 A & 0 &0 & 0 &		 0 & \\{}
 AIII & 0 &0 & 1 & $\mathbb{Z}$ & No ($\mathbb{Z}\to0$)  \\
\rule{0pt}{3ex}  
 AI & $+$ & 0 & 0 & 	0 &  \\
 BDI & $+$ & $+$ & 1 & $\mathbb{Z}$ & Partially yes ($\mathbb{Z}\to\mathbb{Z}_2$) \\
 D & 0 & + & 0 & $\mathbb{Z}_2$ & Yes ($\mathbb{Z}_2\to\mathbb{Z}_2$) \\
 DIII & $-$ & + & 1 & $\mathbb{Z}_2$ & No ($\mathbb{Z}_2\to0$)  \\
 AII & $-$ & 0 & 0 & 0 &  \\
 CII & $-$ & $-$ & 1 & $\mathbb{Z}$ & No ($\mathbb{Z}\to0$)  \\
 C & 0 & $-$ & 0 & 0 &  \\
 CI & $+$ & $-$ & 1 & 0 &  \\
  \hline\hline
 \end{tabular}
  \end{center}
\end{table*}

For a given state, even without knowing the corresponding parent Hamiltonian, entanglement spectrum (ES) \cite{li2008entanglement} is an alternative way to characterizing its topological properties. 
We always prepare a ground state $\ket{\Psi_0}$ of a trivial Hamiltonian $H$ and evolve it with a nontrivial one $H_1$. We consider the case that both the pre-quench and post-quench Hamiltonians belong to the same symmetry class. Evolution of the state is given by $\ket{\Psi(t)}=e^{-i{H_1}t}\ket{\Psi_0}$. At any time $t$, we can divide the system into two parts (for a one-dimensional system, for simplicity, we just divide the system into two halves $L$ and $R$, with equal number of sites), and tracing out the right half to calculate the many-particle ES $\{\lambda_l\}$:
\begin{equation}
	{\rho _L}(t) = {\text{Tr}}_R\left( {\left| {\Psi (t)} \right\rangle \left\langle {\Psi (t)} \right|} \right) = \sum\limits_l {{\lambda _l}(t)} \left| {{\phi _l (t)}} \right\rangle \left\langle {{\phi _l(t)}} \right|,
\end{equation}
where $\left| {{\phi _l}} \right\rangle$ is the basis that diagonalize the reduced density matrix $\rho_L$. 
The single-particle ES $\{\xi_n\}$ and many-particle ES $\{\lambda_l\}$ are related as follows~\cite{fidkowski2010entanglement}:
\begin{equation}\label{Eq:many-particle_ES_single}
	{\lambda _{l = \{ {l_n}\} }} = \prod\limits_n {\left[ {\frac{1}{2} + {l_n}\left( {{\xi _n} - \frac{1}{2}} \right)} \right]} ,\quad {l_n} =  \pm 1.
\end{equation}

We suppose that, for the initial state, as it is a gapped trivial state, there is no single-particle entanglement energy near $1/2$, or equivalently, no degeneracy in the many-particle ES. However, as shown in \cite{gong:2017}, for AZ classes in (1+1)D, there will be  crossings for the time-dependent ES as long as $H_1$ is topologically nontrivial and the energy bands of which are nondispersive (see Table \ref{Tab:classification}). The coincidence of the classifications for (1+1)-dimensional quench dynamics and 1D equilibrium states only happens in the flat-band limit, because the possible time-reversal and chiral symmetries of the parent Hamiltonian will suffer from a {\lq\lq}dynamically induced symmetry breaking{\rq\rq} as indicated by McGinley and Cooper \cite{mcginley2018topology}. When spectrum dispersion of the post-quench Hamiltonian $H_1$ is considered, because of the dynamically induced symmetry breaking, certain symmetry-protected topological classes (AIII, DIII, and CII) will be unstable and become trivial, class BDI is partially stable in the sense that the topological class is reduced (from $\mathbb{Z}$ to $\mathbb{Z}_2$) but still nontrivial, and class D is still stable (see the last column of Table \ref{Tab:classification}, which are the main results of this work). 

It worth mentioning that, the bulk topology for a (1+1)-dimensional system as indicated by the last column of Table \ref{Tab:classification} turns out to be the same as the classification of one-dimensional out-of-equilibrium systems by McGinley and Cooper~\cite{mcginley2018topology}. In Ref.~\cite{mcginley2018topology}, however, the authors use the dynamics of the bulk topological invariant for 1D (i.e., the Zak phase) to diagnosis the (possible) preservation of the nontrivial topology of the initial state. In other word, their work represents a topological classification for a one-dimensional system, rather than for a (1+1)-dimensional system as studied in Ref.~\cite{gong:2017}. 
 
Now that the bulk topology of the (1+1)-dimensional system is generally reduced in the presence of band dispersion, as a consequence of bulk-edge correspondence for the ES~\cite{fidkowski2010entanglement}, ES crossings for the corresponding flat-band topological classes may also be unstable: an entanglement gap could be opened when band dispersion is considered. In the following two sections, we discuss the stability of ES crossings for each flat-band topological classes one by one. We show that, the stable ES crossings in classes D and BDI can be understood as a consequence of parity pump. Using explicit examples in class AIII, we show that ES crossings for the flat-band post-quench Hamiltonian case for classes AIII, DIII, and CII are generally unstable against band dispersion, indicating that the corresponding (1+1)-dimensional bulk topologies are always trivial.

\section{Stable cases}\label{Sec:stable}

After considering the dynamically induced symmetry breaking for the dispersive post-quench Hamiltonian, classes D and BDI remain to be stable and classified by $\mathbb{Z}_2$. The corresponding $\mathbb{Z}_2$ topological invariant is the Fu-Kane invariant~\cite{Fu-Kane2006}, which can be evaluated using the Moore-Balents approach~\cite{moore2007topological}, as detailed in \cite{gong:2017}. We note that, when time-reversal and chiral symmetries are broken dynamically for dispersive post-quench Hamiltonians, the (1+1)-dimensional topology of class BDI is reduced into the one of class D; i.e., these two classes share the same $\mathbb{Z}_2$ topological invariant. As a consequence of bulk-edge correspondence for ES~\cite{fidkowski2010entanglement}, there will be stable ES crossings when the topological number of the post-quench Hamiltonian is odd~\cite{NoteOddES}. 

The robustness of the ES crossings for classes D and BDI can be understood in terms of parity pump for superconducting systems described by the Bogoliubov–de Gennes (BdG) Hamiltonians. For the flat-band BdG Hamiltonians with spectrum $\pm \frac{1}{2}E$, one can show that, after one evolution period $T=2\pi/E$, the parity for the many-particle ES eigenstate $\left| {{\phi _l}(t )} \right\rangle $ changes it sign. To be precise, we consider the initial many-particle ES eigenstate $\left| {{\phi _l}(t=0)} \right\rangle $ as an eigenstate of parity (parity is a conserved quantity for a superconducting system, although the fermion number is not):
\begin{equation}\label{Eq:parity_phi_i}
	{P_L}\left| {{\phi _l}(t = 0)} \right\rangle  = {p_l}\left| {{\phi _l}(t = 0)} \right\rangle ,\quad {p_l} =  \pm 1,
\end{equation}
where 
\begin{equation}
	{P_L} = {( - 1)^{\sum\limits_{i \in L} {{n_i}} }}
\end{equation}
is the parity operator for the left-half system, which counts the parity of the fermion number $n_i$ therein. Then one can show that, the parity for this left-half chain changes sign for each many-body ES eigenstate after one evolution period (see Appendix~\ref{Appendix:A} for details):
\begin{equation}\label{Eq:parity_phi_T}
	{P_L}\left| {{\phi _l}(t = T)} \right\rangle  =  - {p_l}\left| {{\phi _l}(t = T)} \right\rangle .
\end{equation}
The above result holds for all the many-body ES eigenstates. Now we focus on the state with the largest many-body entanglement energy, which is the ground state of the many-body ES. As time goes from $t=0$ to $T$, we see that the parity of this many-particle state changes, which means that at certain time $t_c$ ($0<t_c<T$), the many-body ES ground state and first excited state touches. According to Eq.~\eqref{Eq:many-particle_ES_single}, the single-particle ES also features a crossing at $\xi_n=\frac{1}{2}$, at time $t_c$. By comparing Eqs.~\eqref{Eq:parity_phi_i} and \eqref{Eq:parity_phi_T}, we see that odd fermion number is pumped from the left-half chain to the right-half during the quench dynamics, although the total parity is conserved. 

The above argument of single-particle ES crossing using parity pump relies on the framework of flat-band BdG Hamiltonians. When post-quench Hamiltonian is not flat, we expect that the corresponding ES crossing is still stable against symmetry-preserving perturbations as it is protected by the conservation of the fermion parity~\cite{teo2010topological}. On the other hand, if we consider noninteracting fermion systems (instead of mean-field BdG Hamiltonians) in classes D and BDI, the spectra of them have the same properties as their BdG counterparts at the single-particle level~\cite{Qi2011Review}. Hence even though the ES crossings for such insulating systems are not described by the parity pump, the stability of ES crossings therein can also be inferred from their BdG counterparts. 

In the following, we use concrete models in classes D and BDI to validate the above statements. 

\subsection{Class D}

\begin{figure}[hbtp]
\includegraphics[width=\columnwidth]{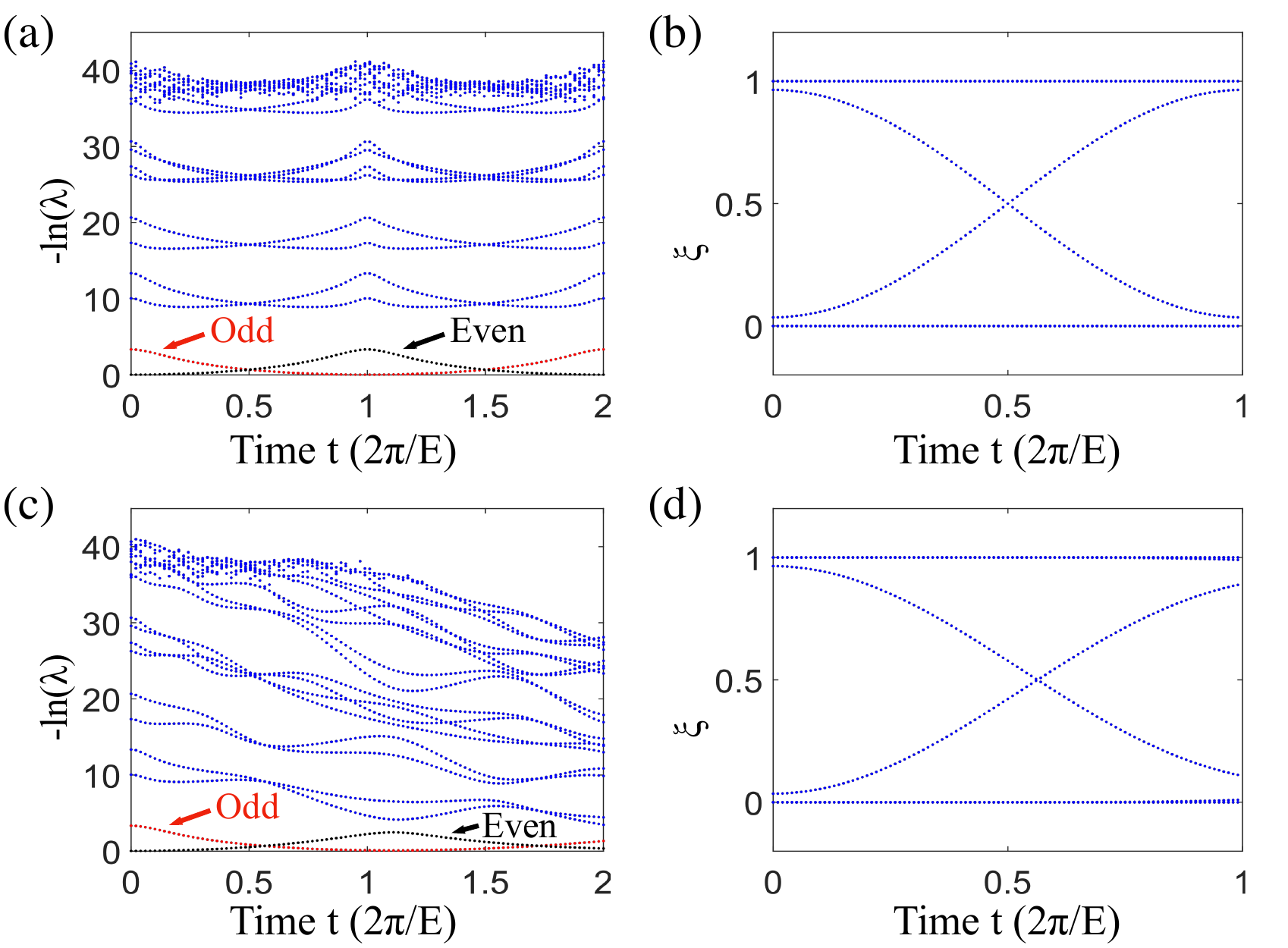}
\caption{Many-particle ES $\lambda$ (a, c) and single-particle ES $\xi$ (b, d) of the quenched Kitaev chain (in class D). In (a, b), we quench the chemical potential $\mu = 1.5\to0$ (the spectrum of the post-quench Hamiltonian is naturally flat for this case); in (c, d), we quench the chemical potential $\mu = 1.5\to0.25$ (the spectrum of the post-quench Hamiltonian is dispersive). The length of the chain is set as $10$. When calculating the ES, the five sites of the right-half chain are traced out. We also explicitly label the parities of the largest two components of the many-particle ES in (a) and (c): red for odd parity and black for even parity. }\label{Fig1}
\end{figure}

First, we consider the Kitaev chain model in class D~\cite{kitaev2001unpaired}:
\begin{equation}
\begin{aligned}
H=\sum_{i} \mu c_i^\dagger c_i + \frac12 \sum_{i}(J c_i^\dagger c_{i+1}+\Delta c_i c_{i+1} + \text{h.c.}),
\end{aligned}
\end{equation}
where $c_i$ is the annihilation operator for a fermion on site $i$, $\mu$ is the chemical potential, $J$ is the nearest-neighbor hopping constant, $\Delta$ is the $p$-wave superconducting pairing amplitude, and h.c. denotes hermitian conjugate. After a Fourier transform into the quasimomentum space, the Hamiltonian takes the following form:
\begin{equation}
	H = \sum\limits_k {\alpha _k^\dag } \left[ {{\mathbf{h}}(k) \cdot \bm{\sigma} } \right]{\alpha _k},
\end{equation}
where $k$ is the quasimomentum, ${\alpha _k} = {({a_k},a_{ - k}^\dag )^T}$ is the Nambu spinor, $\bm{\sigma}=(\sigma_x, \sigma_y, \sigma_z)$ is the vector of Pauli matrices, and ${\mathbf{h}}(k) = (h_x,h_y,h_z) = (0, \frac{1}{2}\Delta \sin k, -\frac{1}{2}(J\cos k + \mu))$. The energy spectrum is given by: $E_\pm(k) = \pm|\mathbf{h}(k)|$. 
We set $J=1$ as the energy unit, and take $\Delta=J$ for simplicity (the value of $\Delta$ does not affect the topology of the system as long as $\Delta\ne0$). When $|\mu|<1$, the system is topological; when $|\mu|>1$, it is trivial. We consider the case that the initial state is trivial, while the post-quench Hamiltonian is nontrivial. The time-dependent many-particle and single-particle ESs are shown in Fig. \ref{Fig1}. In Figs. \ref{Fig1}(a) and (b), we see that, for the quench parameter $\mu = 1.5\to0$, the energy bands of the post-quench Hamiltonian are flat, hence the system features a temporal periodicity $T=2\pi/E$ with $E=J/2$. However, as shown in Fig. \ref{Fig1}(a), the parity of the largest component of the many-body ES [note the minus logarithmic plot of $\lambda$ in Fig. \ref{Fig1}(a), the largest $\lambda$ is the lowest $-\ln(\lambda)$ in the plot] changes from even to odd after one time period, i.e., odd number of fermions are pumped from the left-half chain to the right-half during the time period $T=2\pi/E$. The parity together with the value of the largest many-particle ES component only return to themselves after two evolution periods, which is reminiscent of the fractional Josephson effect~\cite{kitaev2001unpaired,Fu-Kane2009,teo2010topological}. The largest two components of many-particle ES touch each other at time $t=T/2$, and we see that at the same time the corresponding single-particle ES also touches, as shown in Fig.~\ref{Fig1}(b). In Figs. \ref{Fig1}(c) and (d), we show the ES for the quench process $\mu = 1.5\to0.25$, in which the spectrum of the post-quench Hamiltonian is not flat any more. There is also no well-defined temporal periodicity for this case, but we see that the ES crossings for both the many-particle [Fig. \ref{Fig1}(c)] and single-particle [Fig. \ref{Fig1}(d)] cases are preserved, although the touch time is shifted away from $t=\pi/E$. Such crossings are protected by the conservation of fermion parity~\cite{teo2010topological}.

\subsection{Class BDI}

\begin{figure}[btp]
\includegraphics[width=\columnwidth]{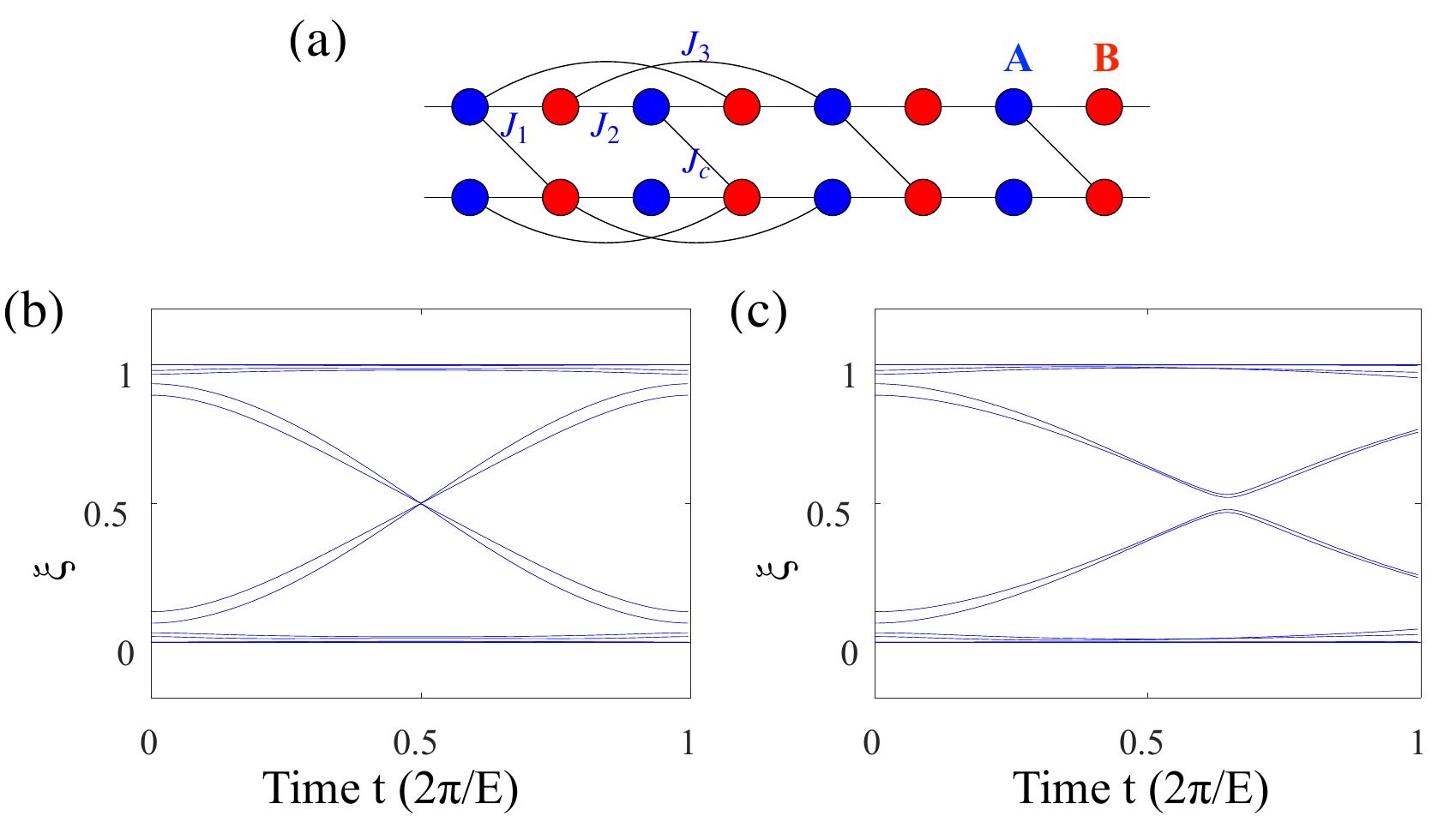}\\
\caption{\label{Fig2} Stability of ES crossing for two coupled SSH-like chains (in class BDI). (a) A schematic illustration of the model, in which the nearest-neighbor ($J_1$ and $J_2$), third-nearest-neighbor ($J_3$), and inter-chain ($J_c$) hopping amplitudes are indicated. This ladder is a bipartite lattice, where the $A$ and $B$ sublattices are denoted in blue and red respectively. (b, c) Evolution of the time-dependent ES for the quench process $(J_1,J_2,J_3,J_c)=(1,0.5,0.6,0.2)E\to(0.5,1,-0.4,0.2)E$. The pre-quench Hamiltonian $H$ is trivial, and the post-quench Hamiltonian $H_1$ is nontrivial with winding number $2$. Here (b) shows the ES calculated with an artificially flattened \cite{gong:2017} post-quench Hamiltonian $\tilde H_1$, while (c) shows the ES with a conventional (non-flat-band) $H_1$. We see that ES crossings in the flat-band case (b) are generally not stable against band dispersion of the post-quench Hamiltonian, and could be gapped (c).}
\end{figure}

Next, we consider concrete examples in class BDI. For this class, if the winding number of $H_1$ is odd, there are always stable ES crossings protected by parity. The behavior of ES is the same as in class D, and we do not repeat the results here for simplicity. However, different from the integer class classification $\mathbb{Z}$ for flat-band post-quench Hamiltonians as shown in Ref.~\cite{gong:2017}, even winding numbers will not promise a crossing when band dispersion is considered. We use two chains of Su-Schrieffer-Heeger (SSH) model~\cite{SSH1979} with additional symmetry-preserving hopping terms, as shown in Fig. \ref{Fig2}(a), to illustrate the triviality for this case. 

The Hamiltonian for this two-leg SSH-like ladder is as follows:
\begin{equation}
\begin{aligned}
H=&\frac12\sum_i [\sum_{j=1,2}(J_1 A_{i, j}^\dagger B_{i, j}+J_2 B_{i,j}^\dagger A_{i+1, j}\\
&+J_3 A_{i, j}^\dagger B_{i+2, j}+J_3 B_{i,j}^\dagger A_{i+2,j})+ J_c A_{i, j=1}^\dagger B_{i,j= 2}+\text{h.c.}]. 
\end{aligned}
\end{equation}
Here, $i$ is the site index (for a unit cell) in the horizontal direction, and $j$ is the index of the leg of the ladder: $j=1$ for the upper leg, $j=2$ for the lower leg. The two sublattices are denoted as $A$ and $B$, and the operators $A_{i,j}$ and $B_{i,j}$ are the annihilation operators for the fermions on the site $(i, j)$ for sublattices $A$ and $B$, respectively. $J_1$ and $J_2$ are nearest-neighbor hopping constants, $J_3$ is the third-nearest-neighbor hopping constant, and $J_c$ is the inter-chain hopping constant. They are all set to be real.

As all the hopping coefficients are real and between sublattices A and B, the model belongs the the class BDI, where the corresponding TRS, CS, and PHS operators are $\Theta=K$, $\Pi=\sigma_z$, and $\Xi=\Theta\Pi=\sigma_z K$, respectively. Here $K$ denotes complex conjugate, and the space of the Pauli matrix $\sigma_z$ is spanned by the two sublattice indexes $A$ (pseudospin up) and $B$ (pseudospin down). There is typically even number of crossings in the ES in the flat-band case, but such crossings are not stable against band dispersion of the post-quench Hamiltonian, see Fig. \ref{Fig2}. The artificial band-flattening process can be found, e.g., in Appendix \ref{Appendix:A} and also in Ref.~\cite{gong:2017}.

\section{Unstable cases}\label{Sec:unstable}

The instability of ES crossings for classes AIII, DIII, and CII against band dispersion can be understood as a direct consequence of dynamically induced symmetry breaking~\cite{mcginley2018topology}. In the following, we use an SSH-like model in class AIII to illustrate such instability. The results for classes DIII and CII can be inferred from the results of class AIII, as the former two classes can be built as a spin-1/2 version of SSH-like model in class AIII; see below for details. 

\subsection{Class AIII}

\begin{figure}[hbtp]
\includegraphics[width=\columnwidth]{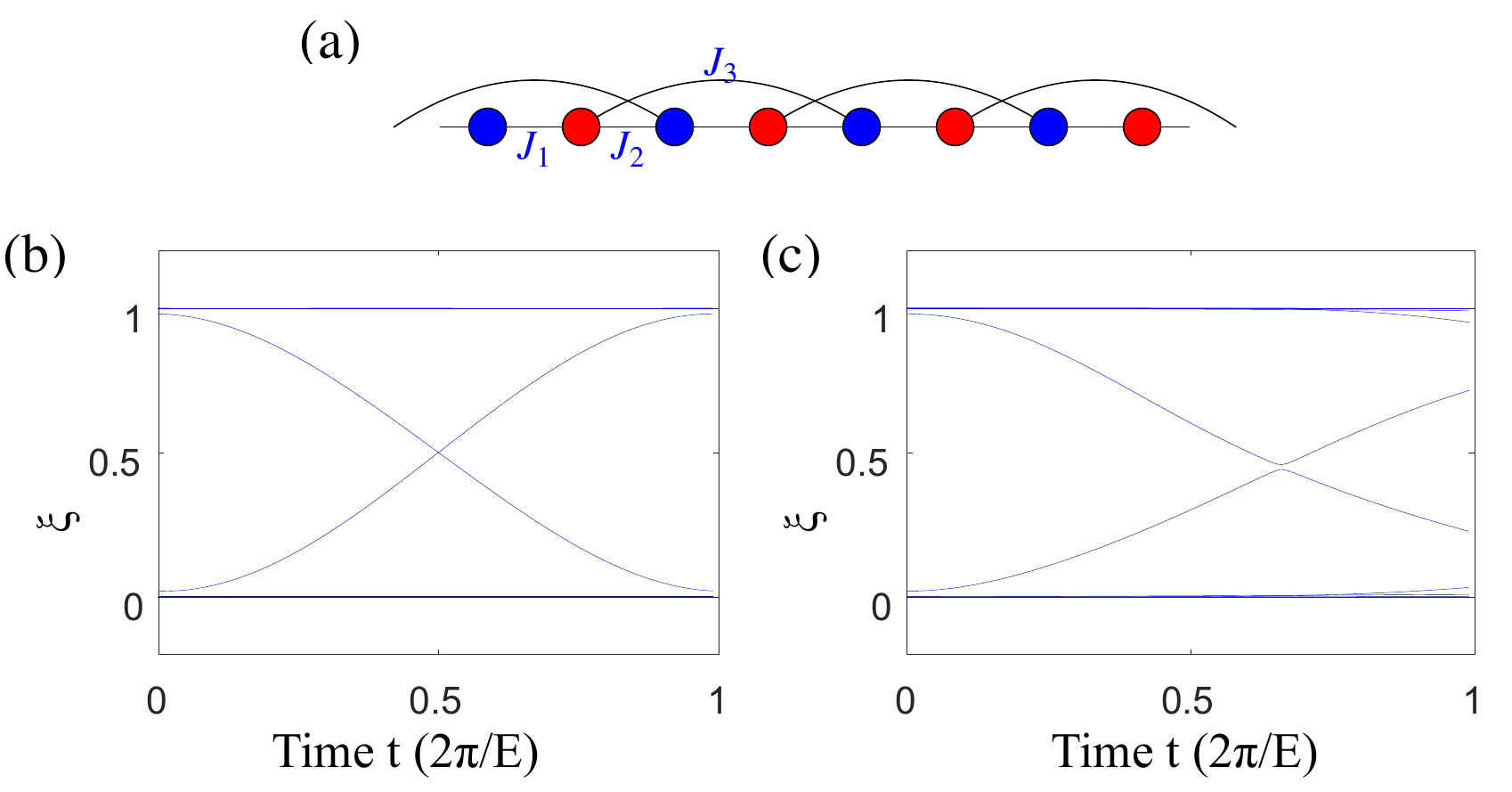}\\
\caption{Stability of ES crossing for an SSH-like model with complex third-nearest-neighbor hopping amplitude $J_3$ (class AIII). (a) A schematic illustration of the SSH-like model, which is just one leg of the SSH-like ladder in Fig.~\ref{Fig2}(a). (b, c) ES evolution of the SSH-like model for the quench process $(J_1,J_2,J_3)=(1,0.5,0.1i)E\to(0.5,1,0.1i)E$. The pre-quench Hamiltonian $H$ is trivial, while the post-quench Hamiltonian $H_1$, which has only chiral symmetry, is nontrivial with winding number $1$. The difference between (b) and (c) is that in (b) we use an artificially flattened post-quench Hamiltonian $\tilde H_1$, while in (c) we use a conventional $H_1$ with a dispersive spectrum (same as the corresponding ones in Fig.~\ref{Fig2}).}\label{Fig3}
\end{figure}

As a concrete model in class AIII, we consider the SSH model with third-nearest-neighbor hopping, as shown schematically in Fig. \ref{Fig3}(a). The Hamiltonian takes the following form: 
\begin{equation}
\begin{aligned}
H=\frac12\sum_i(J_1 A_{i}^\dagger B_{i}+J_2 B_{i}^\dagger A_{i+1}+J_3 B_{i}^\dagger A_{i+2}+\text{h.c.}),
\end{aligned}
\end{equation}
where $i$ is the site index, and $A_i$ ($B_i$) is the annihilation operator for the fermion on the site $i$ for sublattice $A$ ($B$). Here the hopping amplitude $J_1$, $J_2$, and $J_3$ are roughly the same as we have encountered for the SSH-like ladder in class BDI. The only difference is that $J_3$ is taken as a complex number here, which explicitly breaks time-reversal and particle-hole symmetries of the system. But as the hopping is only between $A$ and $B$ sublattices, the chiral symmetry $\Pi=\sigma_z$ is preserved.  When $J_3$ is small enough, whether the system is topologically trivial or not depends on whether $J_2>J_1$ or not. In Figs. \ref{Fig3}(b) and (c), we show the single-particle ES for the quench process from a trivial state evolved by a topological post-quench Hamiltonian for the flat-band and dispersive-band cases. We can see that, the entanglement gap opens when band dispersion is considered. 

\begin{figure}[tb]
\includegraphics[width=0.6\columnwidth]{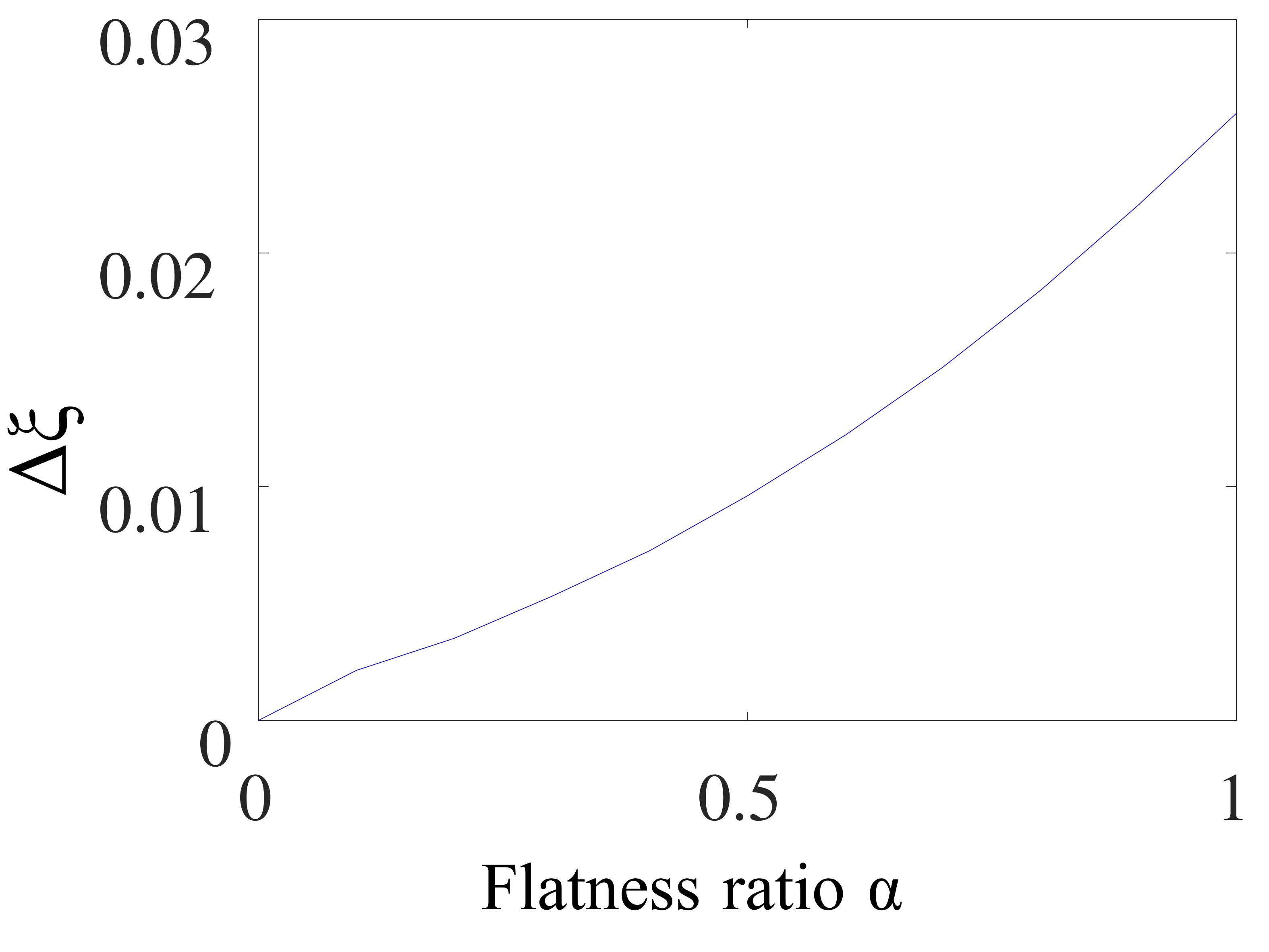}\\
\caption{The entanglement gap $\Delta\xi$ as a function of band flatness ratio $\alpha$ for the quenched SSH-like model as shown in Fig.~\ref{Fig3}. Here $\alpha=0$ correspond to the flat-band limit, for which we see that the entanglement gap vanishes. The entanglement gap opens as long as the band dispersion is considered.}\label{Fig4}
\end{figure}

We can also introduce a band flatness ratio $\alpha$ to show that the crossing is very sensitive to band dispersion. To achieve this, we consider the following set of post-quench Hamiltonians as a function of $\alpha$:
\begin{equation}
	{H_\alpha} = \alpha {H_1} + (1 - \alpha ){{\tilde H}_1}.
\end{equation}
We see that,  $H_{\alpha=0}={\tilde H}_1$ is the artificially flattened post-quench Hamiltonian; $H_{\alpha=1}={ H}_1$ is the conventional, generally dispersive, post-quench Hamiltonian. We consider the same quench process as shown in Fig. \ref{Fig3}, with varying band flatness ratios. The corresponding entanglement gap as a function of the band flatness ratio is shown in Fig.~\ref{Fig4}. We can see that, as long as band dispersion is considered, the entanglement gap opens. This result can be regarded as a direct consequence of the {\lq\lq}dynamically induced symmetry breaking{\rq\rq} \cite{mcginley2018topology} of the parent Hamiltonian. 

\subsection{Classes DIII and CII}

It turns out that the stability of ES crossing in classes DIII and CII can be understood according to the results for class AIII, because they can be considered as two copies of the models that belong to class AIII. We consider a bulk Bloch Hamiltonian in class AIII with the form $H(k)=h_x(k) \sigma_x+h_y(k) \sigma_y$ with chiral symmetry $\sigma_z$. Then  $H^\prime(k)=h_x(k) s_0\otimes\sigma_x +h_y(k) s_0\otimes\sigma_y$ is in class CII with $\Theta=s_y K$, $\Xi= s_y \sigma_z K$, $\Pi=\sigma_z$; and it can also be regarded as a model in class DIII with $\Theta=s_y K$, $\Xi=\sigma_z K$, $\Pi=s_y \sigma_z$. Here $s_0$ is a two-by-two identity matrix and $s_y$ is another Pauli matrix spanned by the real spin (rather than the pseudo-spin as is the case for the model in class AIII) degree-of-freedom. Because the quench dynamics for class AIII is always trivial without stable ES crossing, it is thus also the case for classes DIII and CII as the ES of the latter is just two copies of the former. 

\section{Conclusion}\label{Sec:conclusion}
In conclusion, we have studied the stability of ES crossing for gapped free-fermion systems in (1+1)-dimensional quench dynamics. We find that, unlike the case for the equilibrium system, the band dispersion of the post-quench Hamiltonian reduces the topology of the (1+1)-dimensional system classified by the flat-band parent Hamiltonian, because of dynamically induced symmetry breaking~\cite{mcginley2018topology}. According to the bulk-edge correspondence for the ES, the bulk topology is checked through a thorough investigation of the stability of ES crossing against band dispersion. We find that the stable ES crossing in classes D and BDI can be understood as a parity pump process for BdG Hamiltonians. And the instability of ES crossing for classes AIII, DIII, and CII is illustrated by explicit numerical calculation, with a SSH-like model in class AIII as a concrete example. The stability of ES crossing studied in this work may be validated experimentally through either many-particle state tomography~\cite{Roos2017,Torlai2018}, or direct ES measurement~\cite{Pichler2016,Dalmonte2018}. 

The study of this work can also be extended to quench dynamics in higher dimensions. In particular, we find that the only stable topological class in (2+1)-dimensional quench dynamics is class D with $\mathbb{Z}_2$ topological invariant, which will be elaborated in the future~\cite{Lu2018}.

\begin{acknowledgements}
We would like to thank Xin Chen, Ce Wang, and Hui Zhai for enlightening discussions. This work is supported by MOST under Grant No. 2016YFA0301600 and NSFC Grant No. 11734010. 
\end{acknowledgements}

\appendix
\section{Parity change for many-particle entanglement eigenstates}\label{Appendix:A}
In this appendix, we aim to prove Eq.~\eqref{Eq:parity_phi_T}, which shows that the parity for every many-particle entanglement eigenstate changes sign after time period $T=2\pi/E$ for band flattened BdG Hamiltonians in classes D and BDI. 

We use open boundary condition for the post-quench Hamiltonian $H_1$. As we consider the case that $H_1$ is topological, there are always isolated Majorana zero modes $\gamma_L$ and $\gamma_R$ located on each ends~\cite{kitaev2001unpaired}. The Hamiltonian in the energy basis can be formulated as follows:
\begin{equation}
	{H_1} = \frac{i}{2}\sum\limits_{i = 1}^{N - 1} {{E_i}{\gamma _{2i}}{\gamma _{2i + 1}}}  + 0{\gamma _L}{\gamma _R},
\end{equation}
where $N$ is the total number of lattice sites, $E_i$ denotes the energy of the single-particle bulk state, and $(\gamma _{2i}, \gamma _{2i+1})$ are the corresponding (paired) bulk Majorana operators. We take a band flattening process for the spectrum, and the Hamiltonian takes the following form:
\begin{equation}
	{{\tilde H}_1} = \frac{{iE}}{4}\sum\limits_{i = 1}^{N - 1} {\operatorname{sgn} \left( {{E_i}} \right){\gamma _{2i}}{\gamma _{2i + 1}}} ,
\end{equation}
where $\text{sgn}$ is the sign function, and the spectrum of ${{\tilde H}_1}$ is flattened to be $\pm\frac{E}{2}$. 

The parity for the system can be written as follows:
\begin{equation}
	\begin{aligned}
  P &= {( - 1)^{\sum\limits_{i = 1}^N {{n_i}} }} = {( - 1)^{\frac{i}2\left( {{\gamma _L}{\gamma _R} + \sum\limits_{i = 1}^{N - 1} {{\gamma _{2i}}{\gamma _{2i + 1}}} } \right)+\frac N2}} \hfill \\
   &= {i}{\gamma _L}{\gamma _R}\prod\limits_{i = 1}^{N - 1} {\left( {{i}{\gamma _{2i}}{\gamma _{2i + 1}}} \right)} . \hfill \\ 
\end{aligned} 
\end{equation} 
It is direct to check that 
\begin{equation}\label{Eq:PL_g_PL}
	{P_L}{\gamma _L}{P_L} =  - {\gamma _L},
\end{equation}
which can be seen from 
\begin{equation}
	P{\gamma _L}P = {( - 1)^{2N - 1}}{\gamma _L} =  - {\gamma _L},
\end{equation}
together with $P_R \gamma_L P_R=\gamma_L$. Here we write $P = P_L P_R$, and assume that the isolated Majorana zero mode on the left end has negligible interaction between fermions on the right-half chain. 

The evolution operator for the flat-band Hamiltonian for a period $T=2\pi/E$ is as follows,
\begin{equation}
	U(T) = {e^{ - i{{\tilde H}_1} \cdot 2\pi /E}} = {( - 1)^{{i}\left( {\sum\limits_{i = 1}^{N - 1} {{\gamma _{2i}}{\gamma _{2i + 1}}} } \right)}} = {i}^{1-N}{\gamma _L}{\gamma _R}P. 
\end{equation}
It can be partitioned into two parts as
\begin{equation}
	U(T) = i^{1-N} {U_L}(T){U_R}(T),
\end{equation}
where
\begin{equation}
	{U_L}(T) = {\gamma _L}{P_L},\quad {U_R}(T) = {\gamma _R}{P_R}.
\end{equation}
For the reduced density matrix ${\rho _L}(t) = \sum\limits_l {{\lambda _l}(t)} \left| {{\phi _l}(t)} \right\rangle \left\langle {{\phi _l}(t)} \right|$, we then have 
\begin{equation}
	\begin{aligned}
  {\rho _L}(T) &= \text{Tr}{_R}\left[ {U(T)\rho (0)U{{(T)}^\dag }} \right] \hfill \\
   &= \text{Tr}{_R}\left[ {{U_L}(T)\rho (0){U_L}{{(T)}^\dag }} \right] \hfill \\
   &= \sum\limits_l {{\lambda _l}(0){U_L}(T)\left| {{\phi _l}(0)} \right\rangle \left\langle {{\phi _l}(0)} \right|{U_L}{{(T)}^\dag }}  \hfill \\
   &= \sum\limits_l {{\lambda _l}(0){\gamma _L}\left| {{\phi _l}(0)} \right\rangle \left\langle {{\phi _l}(0)} \right|\gamma _L}  \hfill \\
  &: = \sum\limits_l {{\lambda _l}(0)\left| {{\phi _l}(T)} \right\rangle \left\langle {{\phi _l}(T)} \right|.}  \hfill \\ 
\end{aligned} 
\end{equation}
We thus see that, after time period $T$, each many-particle entanglement energy returns to itself, but the corresponding eigenstate is changed into
\begin{equation}
	\left| {{\phi _l}(T)} \right\rangle  = {\gamma _L}\left| {{\phi _l}(0)} \right\rangle .
\end{equation}
The parity for this state can be calculated as follows:
\begin{equation}
	\begin{aligned}
  {P_L}\left| {{\phi _l}(T)} \right\rangle  &= {P_L}{\gamma _L}\left| {{\phi _l}(0)} \right\rangle  \hfill \\
   &=  - {\gamma _L}{P_L}\left| {{\phi _l}(0)} \right\rangle  =  - {\gamma _L}{p_l}\left| {{\phi _l}(0)} \right\rangle  \hfill \\
   &=  - {p_l}\left| {{\phi _l}(T)} \right\rangle . \hfill \\ 
\end{aligned} 
\end{equation}
Here, Eqs.~\eqref{Eq:PL_g_PL} and \eqref{Eq:parity_phi_i} are used for the second and third equators, respectively. Thus we have proved Eq.~\eqref{Eq:parity_phi_T} as expected. 

% \bibliography{ref}

\begin{thebibliography}{36}
\bibitem [{Not()}]{NoteOddES}%
  \BibitemOpen
  \href@noop {} {}\bibinfo {note} {Note that, we always consider the case the
  the topology of the pre-quench Hamiltonian is trivial. And the topology of
  the (1+1)-dimensional quench dynamics is determined solely by the topology of
  the post-quench Hamiltonian.}\BibitemShut {Stop}%
\end{thebibliography}

\end{document}